\documentclass[twocolumn,showpacs,pre,superscriptaddress,floatfix]{revtex4}

\usepackage{color}
\usepackage{tabularx}
\usepackage{epsfig}
\usepackage{amsmath}
\usepackage{amssymb}
\usepackage{graphicx}
\usepackage{times}

\newcommand{\up}{{\mid \uparrow \rangle}}
\newcommand{\down}{{\mid \downarrow \rangle}}
\newcommand{\LHx}{{{\rm LiHo_xY_{1-x}F_4}}}

\begin{document}

\title{Random-field-induced disordering mechanism in a disordered
ferromagnet:\\ Between the Imry-Ma and the standard disordering
mechanism}

\author{Juan Carlos Andresen}
\affiliation {Department of Theoretical Physics, KTH Stockholm, 10691
Stockholm, Sweden}
\affiliation{Department of Physics, Ben Gurion University of the Negev,
Beer Sheva 84105, Israel}

\author{Helmut G.~Katzgraber}
\affiliation{Department of Physics and Astronomy, Texas A\&M University,
College Station, Texas 77843-4242, USA}
\affiliation{1QB Information Technologies (1QBit), 458-550 Burrard Street,
Vancouver, British Columbia V6C 2B5, Canada}
\affiliation{Santa Fe Institute, 1399 Hyde Park Road, Santa Fe, NM 87501}

\author{Moshe Schechter}
\affiliation{Department of Physics, Ben Gurion University of the Negev,
Beer Sheva 84105, Israel}

\date{\today}

\begin{abstract}

Random fields disorder Ising ferromagnets by aligning single spins in
the direction of the random field in three space dimensions, or by
flipping large ferromagnetic domains at dimensions two and below. While
the former requires random fields of typical magnitude similar to the
interaction strength, the latter Imry-Ma mechanism only requires
infinitesimal random fields. Recently, it has been shown that for dilute
anisotropic dipolar systems a third mechanism exists, where the
ferromagnetic phase is disordered by finite-size glassy domains at a
random field of finite magnitude that is considerably smaller than the
typical interaction strength. Using large-scale Monte Carlo simulations
and zero-temperature numerical approaches, we show that this mechanism
applies to disordered ferromagnets with competing short-range
ferromagnetic and antiferromagnetic interactions, suggesting its
generality in ferromagnetic systems with competing interactions and an
underlying spin-glass phase. A finite-size-scaling analysis of the
magnetization distribution suggests that the transition might be first
order.

\end{abstract}

\maketitle

\section{Introduction}

Spin glasses, where frustration and disorder are introduced through
random competing ferromagnetic and antiferromagnetic interactions
\cite{edwards:75}, and random-field ferromagnets, where disorder is
introduced through an effective longitudinal field \cite{imry:75}, are
two archetypal models for the study of disordered magnetic systems
\cite{binder:86,young:98}. While usually studied independently from each
other, random interactions and random fields are generic in many
nonmagnetic systems \cite{schechter:09}, and dominate thermodynamic
and dynamic properties in e.g. orientational glasses\cite{vugmeister:90}
and relaxor ferroelectrics\cite{westphal:92,sherrington:13}. In magnetic
systems, random interactions are abundant, yet the presence of an
effective longitudinal random field is nontrivial. Applied magnetic
fields cannot be locally randomized and nonmagnetic disorder cannot
produce an effective random magnetic field, because it violates
time-reversal symmetry.

Anisotropic dipolar magnets, and specifically the $\LHx$ compound, are an
intriguing exception. The interplay of an applied field in the direction
transverse to the easy axis of the magnetic holmium ions and the
off-diagonal elements of the dipolar interaction gives rise to an
effective longitudinal field \cite{schechter:06,tabei:06,schechter:08}.
This field is locally random, transforming spatial disorder coming from
the dilution of the Ho ions by the nonmagnetic yttrium ions into a
disorder in the effective longitudinal field. Furthermore, as a function
of holmium concentration, $\LHx$ has both a ferromagnetic phase at $x
\gtrsim 0.3$ and a spin-glass phase at $0 < x \lesssim 0.3$ including
the extreme dilute limit \cite{tam:09,andresen:13b,andresen:14}. The
$\LHx$ system is therefore ideal for the study of the interplay of
competing interactions and effective longitudinal random fields, and the
effect of this interplay on the thermodynamic and dynamical properties
of the system.

Recently, we have shown \cite{andresen:13b} that disordered anisotropic
dipolar magnets in their ferromagnetic phase, are driven into a
quasi-spin-glass phase upon the introduction of a {\it finite} effective
random magnetic field which is {\it considerably smaller than the
typical nearest-neighbor interaction}. This occurs also in three space
dimensions, where it displays a novel disordering mechanism,
intermediate between the standard disordering of a ferromagnet and the
Imry-Ma \cite{imry:75} mechanism at dimensions two and below. The
disordering field is neither infinitesimal, nor of the order of the
interaction. The smallness of the random field at the transition is
dictated by the proximity of the system to the zero-field transition
between the ferromagnet and the spin-glass phase. Moreover, the
disordered phase near the transition consists of neither single spins
pointing in the direction of the random field, nor of large
ferromagnetic domains, but of finite-size glassy domains, reminiscent of
the competing spin-glass phase. We denote this phase
``quasi-spin-glass'' (QSG) in the regime of low temperatures, where it
is frozen, and ``paramagnetic QSG'' (pQSG) in the high temperature
regime. The size of the glassy domains is a function of the magnitude of
the disordering field and temperature
\cite{schechter:06,andresen:13b,comment:aharony}.

Here we consider a more general model of an Ising magnet where the
interactions are short ranged, taken from a Gaussian distribution with a
mean ferromagnetic value $J_0$ and a standard deviation $J$, and with
longitudinal random fields taken from a Gaussian distribution of mean
zero and standard deviation $H_r$. Using jaded extremal optimization
\cite{middleton:04} and large-scale parallel tempering Monte Carlo
simulations \cite{hukushima:96} we analyze the thermodynamic properties
of the system at zero and finite temperature. For zero random field we
establish the phase diagram of the system, consisting of low-temperature
ferromagnetic (FM) and spin-glass (SG) phases and a high-temperature
paramagnetic (PM) phase. The zero-temperature transition between the FM and
SG phases occurs at a ratio of $J/J_0 \approx 1.65$. We then
analyze the disordering of the FM phase at $J/J_0 < 1.65$
with increasing temperature and random field. For zero temperature we
find that the disordering of the FM phase occurs at
finite random field, which is much smaller than the typical interaction,
$0 < H_r \ll J_0$, the value of $H_r$ depending on the proximity of the
system at zero field to the SG phase.  At finite
temperature we find that the critical temperature of the
FM-to-pQSG transition increases linearly as a
function of decreasing random field, down to a rather small value of
$H_r$. Our results here are in agreement with our previous results for
the dipolar-interacting $\LHx$ system \cite{andresen:13b}, as well as
with experimental data for this material \cite{silevitch:07}. This
suggests that the disordering mechanism by finite-size glassy domains is
a {\em generic} feature of disordered ferromagnets with competing
interactions in the presence of random fields. By analyzing the
distribution of magnetization values near the random-field-driven
transition, we find evidence for a first-order transition between the FM
and pQSG phases in the regime where the FM phase is in proximity to the SG
phase. No evidence for a first-order transition is found in the regime
where the interactions are strongly ferromagnetic dominated, i.e., $J/J_0
\ll 1$.

The paper is structured as follows. In Sec.~\ref{sec:Model} we introduce
the model. In Sec.~\ref{sec:Methods} we introduce the methods used for
zero-temperature and for finite-temperature calculations.  Our results
are presented in Sec.~\ref{sec:Results}, and discussion and conclusions
are given in Sec.~\ref{sec:Discussion}. An Appendix lists all parameters
of the different simulations.

\section{Model}
\label{sec:Model}

The model we simulate is given by the Hamiltonian
\begin{align}
\mathcal{H} &= - \sum_{\left< i,j\right>} J_{ij} S_i S_j - \sum_i h_i S_i\,,
\label{eq:hamiltonian}
\end{align}
where the sum is over nearest neighbors, the spin couplings $J_{ij}$ are
chosen from a Gaussian distribution with standard deviation $J$ and mean
$J_0$, and the $S_i$ are Ising spins on the vertices of a hypercube of
dimension $d=3$ and linear size $L$. Throughout this work we fix the
mean of the distribution $J_0=1$ and use the standard deviation of the
Gaussian distribution $J$ as a means of tuning the interaction disorder
in the model. The second term describes the coupling of the Ising spins
to a site-dependent random field $h_i$. The random fields are taken from
a Gaussian distribution with zero mean and standard deviation $H_r$.

\begin{figure}
\includegraphics[width=0.8\columnwidth]{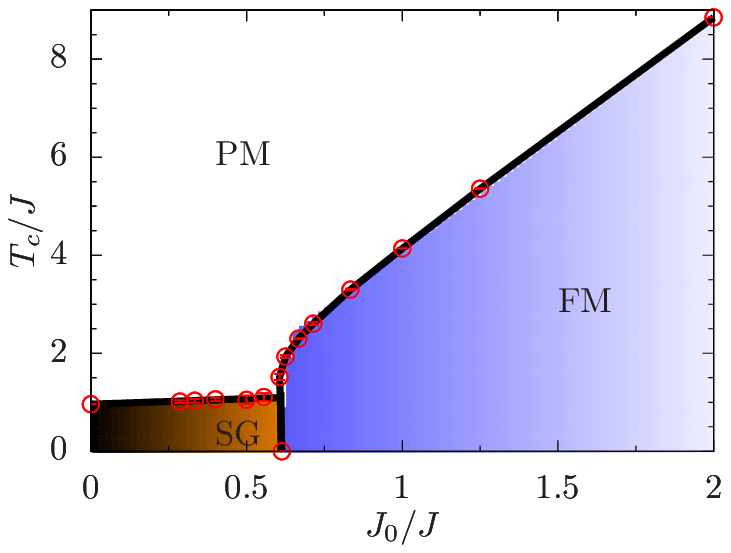}
\caption{
Dimensionless temperature $T/J$ vs mean of the disorder distribution
$J_0/J$ phase diagram for the model given in Eq.~(\ref{eq:hamiltonian})
with fixed $J_0=1$ and $H_r=0$. At $J\lesssim 1.65$
($J_0/J\gtrsim0.606$) and low temperatures the system is in a FM phase.
For $J\gtrsim 1.65$ ($J_0/J\lesssim0.606$) and low temperatures the
system is in a SG phase.  At high temperatures the system is in the PM
phase independently of the value of $J_0/J$.  Note that the model
reduces to the three-dimensional Ising model for $J=0$ and to the EA
spin-glass for $J_0=0$.}
\label{fig:tj_phase_diagram}
\end{figure}

Different parameters of the model lead to different phase diagrams. If
$J$ is small compared to $|{J_0}|$ and $H_r=0$, as the temperature is
reduced, the system undergoes a phase transition from a PM phase to a FM
phase for $J_0>0$. For $J_0<0$ the system undergoes a phase transition
from a PM phase to an antiferromagnetic (AFM) phase.  By increasing the
value of the parameter $J$ it is possible to introduce disorder and
frustration into the system. These two ingredients are known to be
essential for the emergence of a SG phase. The ratio $J_0/J$ quantifies
the amount of interaction disorder in the system. Below a critical ratio
$J_0/J_c$, the system has a PM phase at high temperatures and a SG phase
at low temperatures (keeping $H_r=0$). Extreme cases occur when $J=0$,
$J_0\neq0$ ($J_0/J\to\infty$) where the model reduces to the well-known
Ising model if $H_r=0$, and to the random-field Ising model
\cite{imry:75} if $H_r>0$; and when $J\ne0$, $J_0=0$ ($J_0/J=0$) where
the model reduces to the Edwards-Anderson (EA) spin-glass model
\cite{edwards:75}. In Fig.~\ref{fig:tj_phase_diagram} we plot the phase
diagram of the system as a function of temperature and interaction
disorder $J$ (keeping $H_r=0$) presenting a phase transition at zero
temperature between a FM phase at small disorder and a SG phase at large
disorder, as well as a PM phase at high temperatures.

The random-field term in Eq.~(\ref{eq:hamiltonian}) introduces a third
axis to Fig.~\ref{fig:tj_phase_diagram}. For both limits of zero $J$ and
finite $J_0$ (pure ferromagnet) and zero $J_0$ and finite $J$ (pure EA
spin glass) the effect of finite random field has been thoroughly
studied \cite{young:85, rieger:93, rieger:95, ogielski:86,
young:98,sethna:93,perkovic:95, bertotti:06, thouless:80, fisher:86,
fisher:87, katzgraber:05c}. For the latter limit two different
pictures to describe finite-dimensional spin glasses have been proposed:
the replica symmetry breaking (RSB) picture based on the solution of the
Sherrington-Kirkpatrick
model~\cite{parisi:79,parisi:83,rammal:86,mezard:87,parisi:08}, which
predicts the existence of a spin-glass phase at finite fields and the
droplet picture~\cite{mcmillan:84,fisher:86,fisher:87,fisher:88}, which
contrary to the RSB predicts the instability of the spin-glass phase for
any infinitesimal small field. Which of these pictures correctly
describes the three-dimensional spin glass is still an open
question~\cite{young:04,katzgraber:05c,joerg:08a,katzgraber:09b,fernandez:10,yucesoy:12,billoire:14a,morais:16}.
In this paper, our theoretical considerations are based on the droplet
picture, but we do not undermine the possibility to derive an analogous
prediction within the RSB picture. Specifically, within the droplet
picture of spin glasses, it is argued \cite{fisher:86} that the
spin-glass phase is unstable to infinitesimal random fields, as
finite-size glassy domains are created, destroying long-range glass
order. Here we study the effect of the random field in the range, where
the system is a ferromagnet, albeit with competing interactions
($0<J/J_0<1.65$). At $T=0$ we obtain the phase diagram as a function of
$J$ and $H_r$. At finite temperature we obtain the phase diagram for
different values of $J$ in the ferromagnetic regime, in proximity to the
spin-glass phase and deep inside the ferromagnetic regime. We find the
dependence of the critical temperature $T_c$ on random-field strength
and study the nature of the phase transition.

\section{Methods}
\label{sec:Methods}

\subsection{Zero-temperature simulations}

For the zero-temperature simulations we use the jaded extremal
optimization heuristic \cite{boettcher:01,middleton:04}. We set the
algorithm parameter $\tau=1.6$, $1.8$, $1.9$ with an aging parameter
$\Gamma = 0.0001$ for at least $2^{24}$ simulation steps. Each disorder
realization undergoes at least $512$ independent runs.  We monitor how
many times the lowest energy configuration is found. When the success
rate is more than $\sim 5\%$, we assume to have found the ground-state
configuration. Ground states are found with high confidence for
$L\leqslant 10$ for $H_r=0$ and $L\leqslant 8$ for $H_r\neq 0$.

The FM to SG (PM) phase boundary is identified through the Binder
ratio \cite{binder:81}
\begin{align}
g = \frac{1}{2}
    \left(
     	3 - \frac{\left[m^4\right]_{\rm av}}{\left[m^2\right]_{\rm av}^2}
    \right) ,
\label{eq:binder}
\end{align}
where $m=1/N \sum_iS_i$ is the magnetization of the system, $N=L^3$ is
the number of spins and $\left[ \cdots \right]_{\rm av}$ represents a
disorder average. The Binder ratio $g$ is a dimensionless observable
that scales as $g\sim \tilde{G} \left[ L^{ 1/\nu}(J-J_c)\right]$. The
argument vanishes if $J=J_c$ for all linear system sizes $L$. Therefore,
the crossing of the curves for different system sizes gives an estimate
of the transition value $J_c$ up to finite-size effects. Simulation
parameters are shown in table~\ref{table:04}.

\begin{figure}[h]
\includegraphics{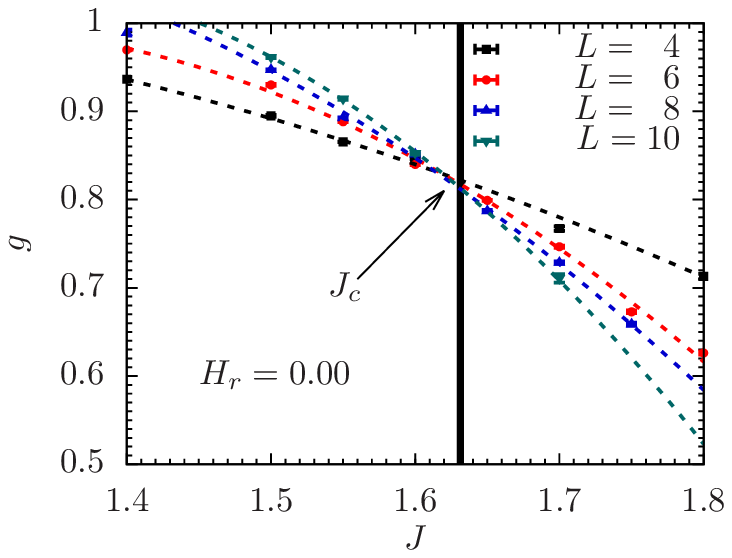}
\caption{
Binder ratio $g$ as given by Eq.~(\ref{eq:binder}) for linear system
sizes $L=4$, $6$, $8$, and $10$ at a random field strength $H_r=0$ and
$T=0$. The crossing of the curves for different sizes $L$ gives an
estimate of the critical disorder $J_c$ denoting the transition between
the FM phase at $J<J_c$ and the SG phase at $J>J_c$. The dashed lines
are guides to the eye.
\label{fig:binder}}
\end{figure}

The critical disorder $J_c$ and the standard deviation for each random
field $H_r$ are estimated using a Levenberg-Marquardt minimization
combined with a bootstrap analysis, where we assume that the universal
function $g$ is well approximated by a third-order polynomial. We show
in Figure~\ref{fig:binder} how the estimated $J_c$ agrees with the
crossing of the Binder ratio curves when $H=0.00$.  The dashed lines are
fit functions to the data used only for visualization purposes and the
vertical line is the average of the intersections of all samples, in
this case $J_c=1.63(1)$.

\subsection{Finite-temperature simulations}

The simulations at finite temperature are done using a combination of
single-spin-flip Monte Carlo with parallel tempering Monte Carlo
\cite{hukushima:96}. To determine the finite-temperature transitions for
a fixed random field strength $H_r$ and disorder $J$ we measure the
ferromagnetic and spin-glass two-point correlation length
\cite{ballesteros:00}
\begin{align}
\xi_{L}^{\rm sg,fm} = \frac{1}{2\sin \left(k_{\rm min}/2\right)}
\sqrt{\frac{\chi_{\rm sg,fm}({\bf 0})}{
\chi_{\rm sg,fm}({\bf k}_{\rm min})} -1}\,,
\label{eq:correlation}
\end{align}
where ${\bf k}_{\rm min} = (2\pi/L,0,0)$, $\chi_{\rm sg}\left({\bf
k}\right)$ is the spin-glass wave-dependent susceptibility
\begin{align}
 \chi_{\rm sg}\left({\bf k}\right)
= \frac{1}{N}\Big[\langle\sum_{i,j}
 q_iq_je^{\text{i}{\bf k}
  {\bf r}_{ij}}
 \rangle\Big]_{\rm av}\,,
\end{align}
$q_i=S_i^\alpha S_i^\beta$ is the two-replica spin overlap,
and $\chi_{\rm fm}\left({\bf k}\right)$ is the ferromagnetic wave-dependent susceptibility
\begin{align}
 \chi_{\rm fm}\left({\bf k}\right)
= \frac{1}{N}\Big[\langle \sum_{i,j}
 S_iS_j e^{\text{i}{\bf k}
  {\bf r}_{ij}}\rangle\Big]_{\rm av}\,.
\end{align}
Here, $\langle \cdots \rangle$ denotes a thermal average and
$\left[\cdots\right]_{\rm av}$ a disordered average.

\begin{figure}
\includegraphics[width=0.8\columnwidth]{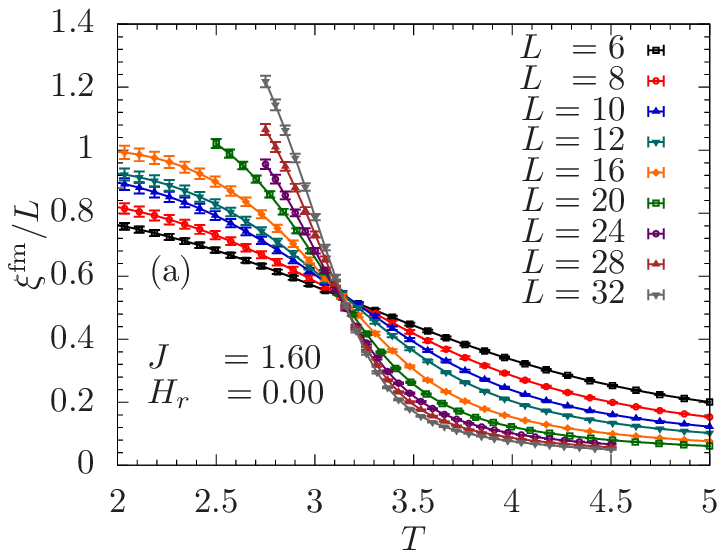}
\includegraphics[width=0.8\columnwidth]{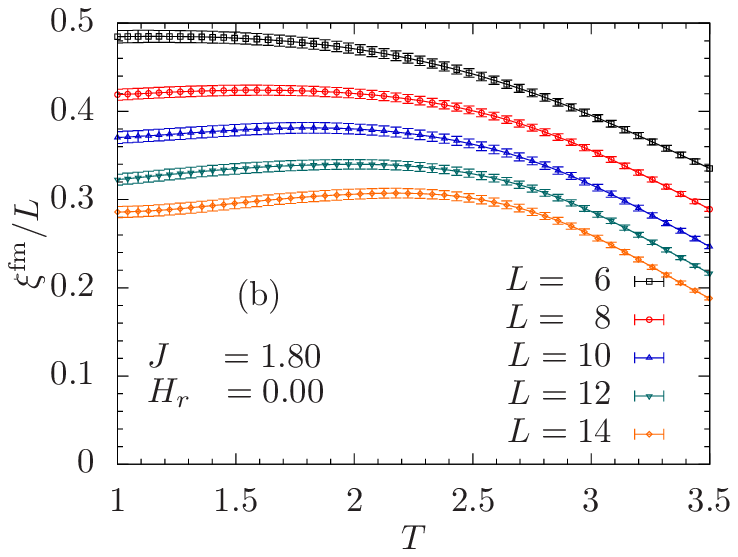}
\includegraphics[width=0.8\columnwidth]{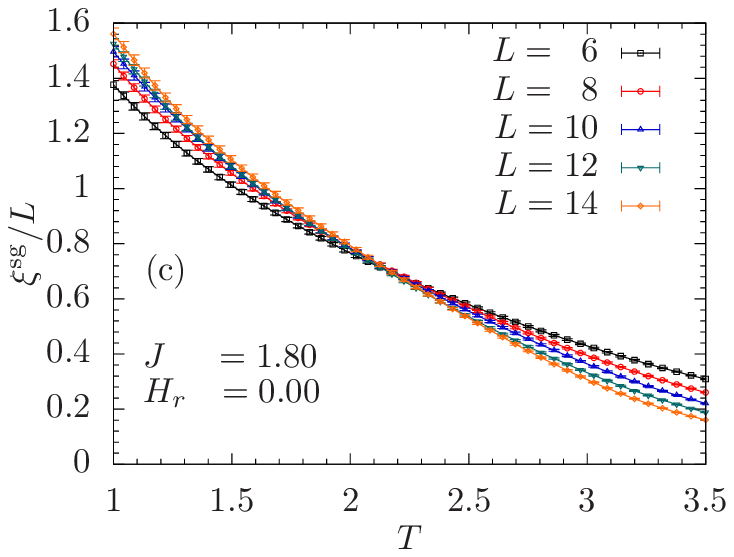}
\caption{
Ferromagnetic and spin-glass two-point correlation function divided by
the linear system size $L$ for $J=1.60$ [panel (a)] and $1.80$ [panels
(b) and (c)], respectively. (a) The crossing of the curves of the
dimensionless quantity $\xi^{\rm fm}_L/L$ for linear system sizes $L=6$
-- $20$ indicates that a PM to FM transition occurs at $T_c\sim3.09(4)$.
(b) The dimensionless quantity $\xi^{\rm fm}_L/L$ does not cross in the
studied temperature range, showing the lack of a FM transition. (c)
Curves corresponding to different system sizes of the dimensionless
quantity $\xi^{\rm sg}_L/L$ cross at $T_c\sim1.90(14)$ indicating that a
PM to SG transition occurs in this temperature range.
\label{fig:xl}}
\end{figure}

Near the transition the dimensionless ratio of the two-point correlation
functions $\xi_L^{\rm sg,fm}$ and the linear system size $L$ scales as
$\xi_L^{\rm sg,fm}/L \sim \tilde X \left[L^{1/\nu} \left(T-T_c\right)
\right]$. At the critical temperature the argument vanishes and the
dimensionless quantity becomes size independent (up to scaling
corrections), hence we expect lines of different system sizes to cross
at $T_c$. If, however, the lines do not cross, no transition takes place
at the studied temperature range. Some example cases are shown in
Fig.~\ref{fig:xl}. In Fig.~\ref{fig:xl}(a) the FM two-point correlation
length for $J=1.60$ ($J_0/J=0.625$) and $H_r=0.00$ is depicted. Clearly,
the curves cross at a putative point which indicates a transition at a
temperature $T_c$. In Fig.~\ref{fig:xl}(b) we plot the same quantity,
but for a higher disorder value $J=1.80$ ($J_0/J=0.5\bar{5}$). In this
case, clearly the curves do not intersect at any studied temperature,
showing a lack of a ferromagnetic phase transition in this temperature
range. Moreover, it suggests a possible lack of ferromagnetic transition
at any $T>0$. In Fig.~\ref{fig:xl}(c) we show the spin-glass two-point
correlation length for the same disorder value $J=1.80$ as in
Fig.~\ref{fig:xl}(b). The curves do cross in the simulated temperature
range, signaling a PM to SG phase transition.

For small random field strengths and interactions distribution widths
the equilibration time is relatively short, making it possible to
simulate large system sizes. With increasing random fields and
interaction distribution widths equilibration times become longer.  To
determine the critical temperature we approximate the scaling function
by a third-order polynomial and perform a fit with six free parameters.
To estimate the error bars we use a bootstrap analysis as described by
Katzgraber {\em et al.}~\cite{katzgraber:06}. Simulation parameters for
the finite-temperature simulations are shown in Tables~\ref{table:01},
\ref{table:02}, and \ref{table:03}.

\begin{figure}
\includegraphics[width=0.8\columnwidth]{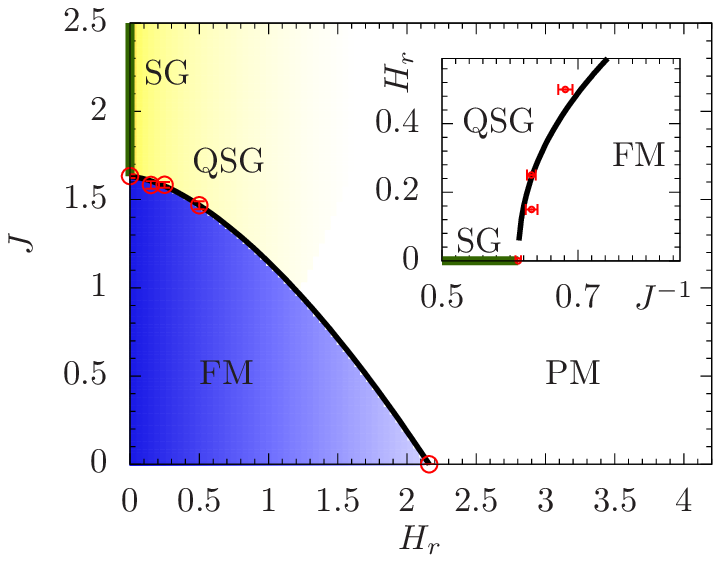}
\caption{
$J$--$H_r$ phase diagram at $T=0$. At $H_r=0$ there is a FM-SG
transition at $J_c\left(0\right)=1.63$. At finite but small $H_r$
the disordered paramagnetic phase has finite size glassy domains,
denoted here as a ``quasi-SG" (QSG) phase. The crossover between the
QSG phase and the PM phase with short range spin-spin correlations is
speculative. In the inset, the solid black line is
a fit $H_r\left(J^{-1}\right)=\alpha\,(J^{-1}-J_c^{-1})^\beta$ to the
data at low random field strength.  Here, $J_c$ and $\beta$ are fixed
parameters; we use the above estimated value $J_c\left(0\right)=1.63$,
and the analytical result of Ref.~\cite{andresen:13b} $\beta=0.466$.
$\alpha = 1.52(17)$ is a free fitting parameter.
\label{fig:phase_diagram_h_sigma}}
\end{figure}

\begin{figure*}
\includegraphics[width=0.32\textwidth]{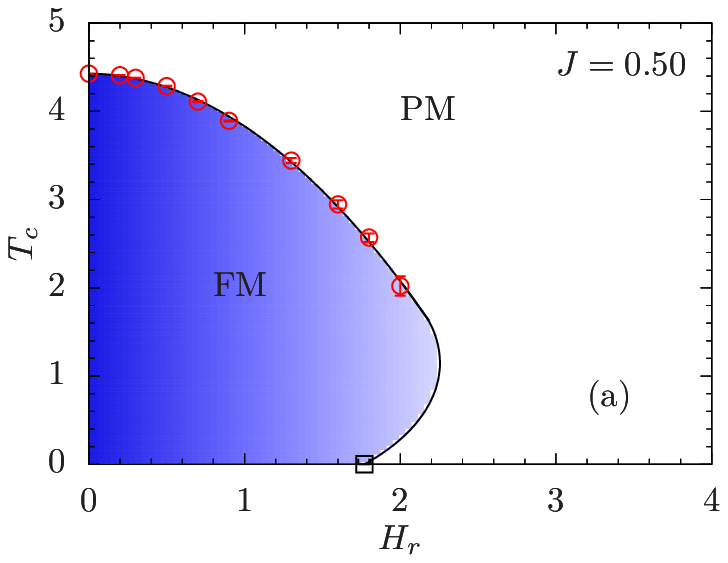}
\includegraphics[width=0.32\textwidth]{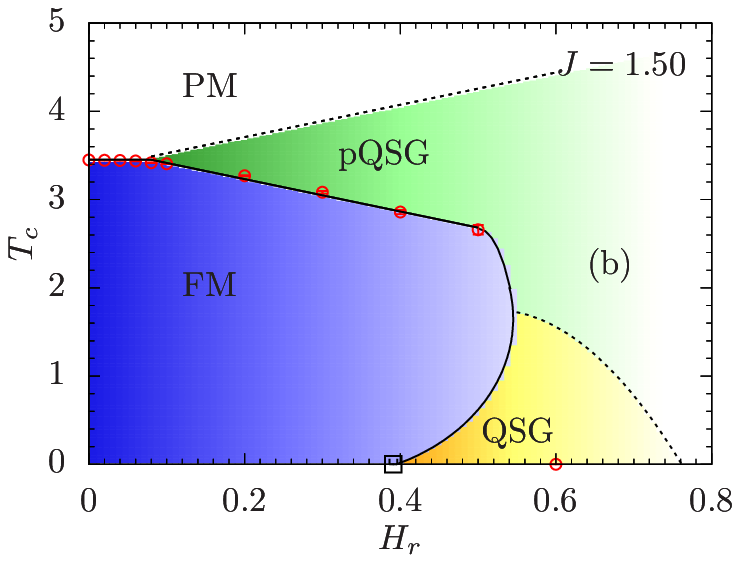}
\includegraphics[width=0.32\textwidth]{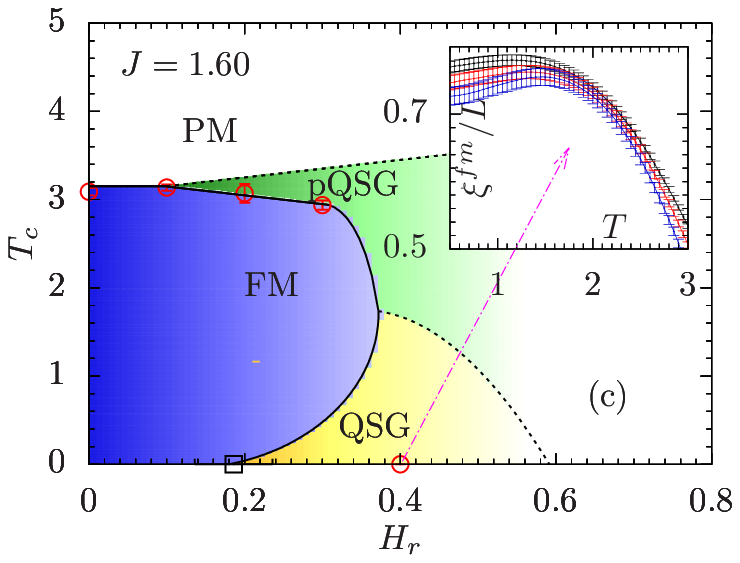}
\caption{
$T$--$H_r$ phase diagram for $J=0.50$ [panel (a)], $J=1.50$ [panel (b)]
and $J=1.60$ [panel (c)]. For the slightly-disordered system with
$J=0.50$ in panel (a) the relation expected from mean-field theory
$T\left(H_r\right)-T_c \sim H_r^2$ holds, except for a weak reentrance
at the lowest temperatures.  The data are consistent with the disordered
phase being a standard PM. For the highly-disordered systems with
$J=1.50$ [panel (b)] and $J=1.60$ [panel (c)] where the FM phase is
close to the SG phase at $H_r=0$ (see
Figure~\ref{fig:phase_diagram_h_sigma}) we find a linear relation
$T\left(H_r\right)-T_c(h^*) \propto H_r$ for $H_r>h^*$ consistent with
the disordered phase being a pQSG. Furthermore, we find strong
reentrance at low temperatures with critical $H_r \ll J_0$, in agreement
with the disordered phase being a QSG.  The inset in panel (c) shows
$\xi^{\rm fm}_L$ for $L=8$, $10$, and $12$ for $H_r=0.40$ where the
curves do not cross, indicating the lack of a phase transition to the FM
phase for the studied temperature range. The dashed lines in all panels
between the pQSG and PM phases and between the QSG and pQSG phases
denote smooth crossovers, their functional form should be considered as
a guide to the eye.
\label{fig:th_phase_diagram}}
\end{figure*}

\begin{figure*}
\includegraphics[width=0.32\textwidth]{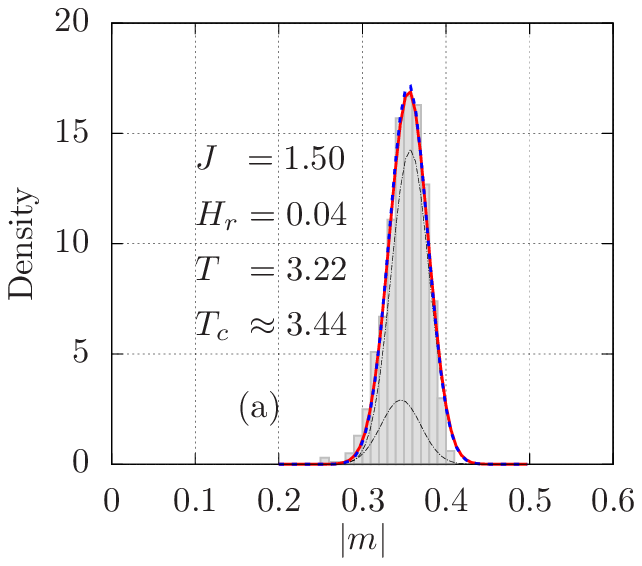}
\includegraphics[width=0.32\textwidth]{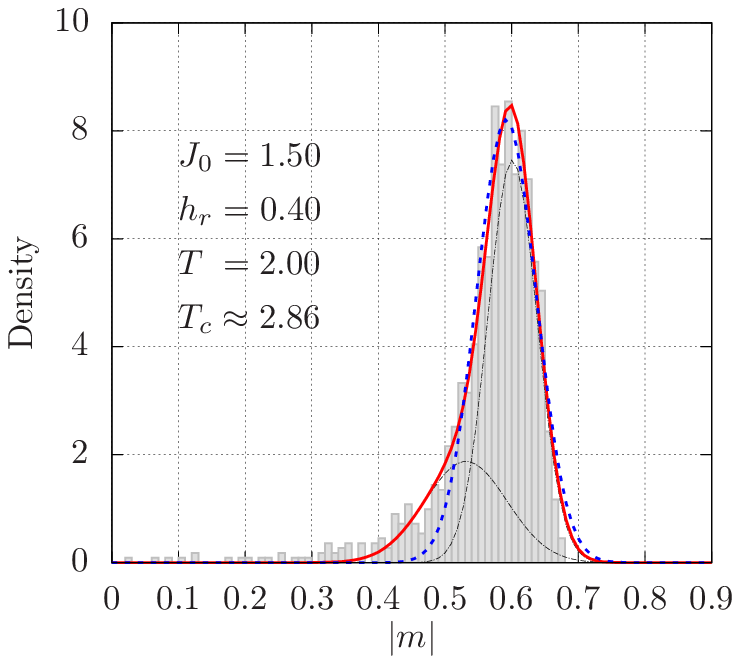}
\includegraphics[width=0.32\textwidth]{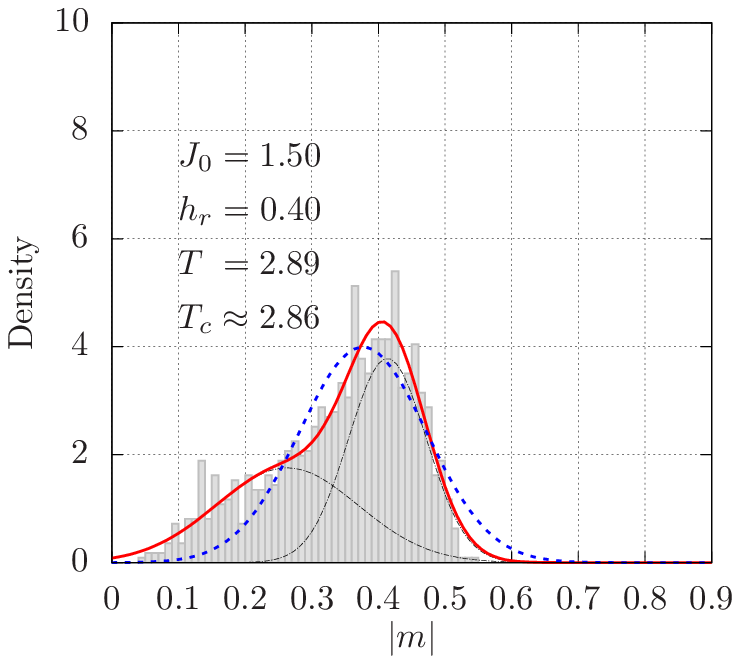}
\caption{
Magnetization histograms for $J=1.50$ with a random field $H_r=0.04$ and
$L=40$ (below the critical temperature) [panel (a)], $H_r=0.40$ and
$L=20$ (below the critical temperature) [panel (b)], and $H_r=0.40$ and
$L=20$ (above the critical temperature) [panel (c)].  The solid red
lines are double-Gaussian fits, which are composed of the sum of the two
modes (thin dashed black lines), the blue dashed lines are single
Gaussian fits. For $H_r=0.04$, the double Gaussian fit does not differ
much from the single Gaussian fit, i.e., the distribution is normally
distributed as expected for all temperatures in a continuous phase
transition. For $H_r=0.40$ a hump emerges for $T<T_c$, the single
Gaussian fit (blue dashed line) cannot capture the hump, but the double
Gaussian fit (solid red line) does.  Close to the critical temperature
the single Gaussian fails to fit the distribution and a bimodal
double-Gaussian structure becomes evident suggesting a finite jump in
the magnetization washed out by the disorder at $T_c$, as expected for a
first-order phase transition.
\label{fig:magnetization_histogram}}
\end{figure*}

\begin{figure*}
\includegraphics[width=0.32\textwidth]{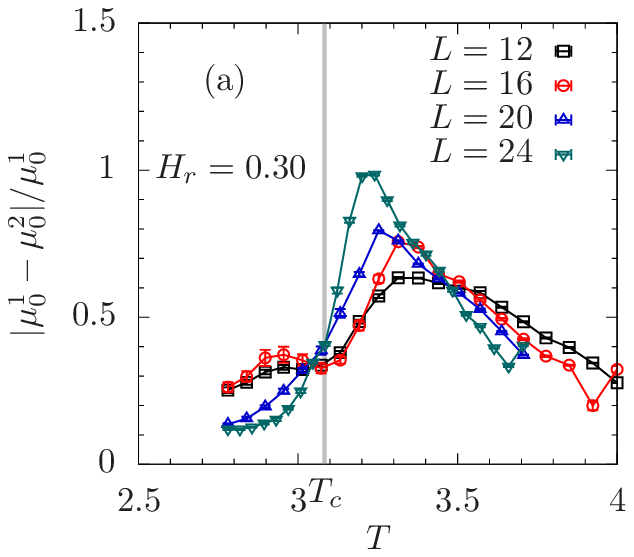}
\includegraphics[width=0.32\textwidth]{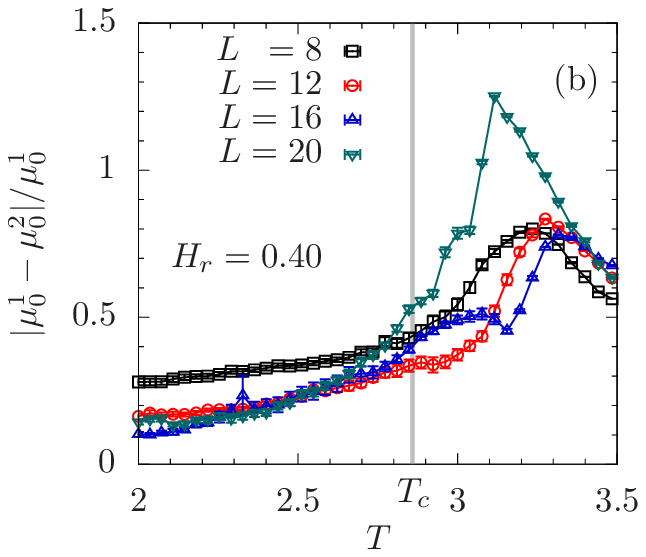}
\includegraphics[width=0.32\textwidth]{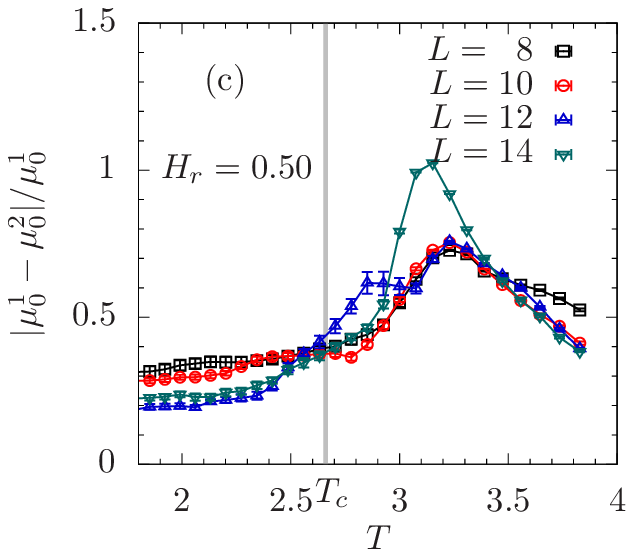}
\caption{
Relative difference of the bootstrapped mean value of the two modes
(with mean values $\mu_0^1$ and $\mu_0^2$) of the bimodal Gaussian
fitted function as a function of temperature for (a) $H_r=0.30$, (b)
$H_r=0.40$, and (c) $H_r=0.50$. The gray vertical line corresponds to the
estimated critical temperature. The relative difference between the
modes increases close to $T_c$, the discrepancy between the
estimated value for $T_c$ and the maximal values of the curves can be
attributed to the strong finite-size correction of this observable.
\label{fig:mean_histogram}}
\end{figure*}

\section{Results}
\label{sec:Results}

The phase diagram of the model presented in Sec.~\ref{sec:Model} is
shown in Fig.~\ref{fig:tj_phase_diagram}. It has a high temperature PM
phase and low temperature FM and SG phases at small and strong disorder
$J$, respectively. The boundary between the FM and the SG phase is at
$J\approx1.65$ ($J_0/J\approx0.606$) with a weak reentrance at $T=0$
[$J_c=1.63(1)$].  The obtained phase diagram qualitatively agrees with
previous results\cite{sherrington:75a,sherrington:75,southern:77} for
the model in Eq. (\ref{eq:hamiltonian}), and for the closely related
diluted bimodal Ising SG model\cite{hasenbusch:07a}. In comparison to
the real space rescaling method by Southern and Young
\cite{southern:77}, we find larger values for $T_c$ and for $J_c$, in
accordance with the typical behavior of real-space rescaling method
\cite{southern:77}. Note the linear dependence of the PM to SG
transition on the mean interaction value. The same is true for the PM to
FM phase transition in the regime $J\lesssim 1.2$ ($J_0/J\gtrsim
0.83\bar{3}$).

We now study the phase diagram with the inclusion of finite random
fields and consider first the case where $T=0$, see
Fig.~\ref{fig:phase_diagram_h_sigma}. The FM phase at low disorder $J$
and low random-field strength $H_r$ is disordered at large $J$ into a
QSG phase, consisting of finite size glassy domains, but no long range
glass order except at $H_r=0$. The data point at $H_r=0$ corresponds to
the FM to SG transition point at $T=0$ plotted also in the $T/J$ vs
$J_0/J$ phase diagram shown in Fig.~\ref{fig:tj_phase_diagram}.  At
small $J$ and large $H_r$ the FM phase disorders into a PM phase. The
data point at $J=0$ corresponds to the critical random field value $H_r
\approx 2.16$ of the random-field Ising model. For low disorder $J$ we
expect the critical random field to be of the order of the mean
interaction $J_0$, because in the PM phase single spins order along
their local effective field.

The situation for $J \lesssim J_c$ is, however, different. At $H_r=0$
the FM ground state and the lowest-energy SG state differ in energy, the
difference being linear in $J-J_c$ \cite{andresen:13b}. At finite $H_r$,
the low-energy SG state changes profoundly: long-range order is
destroyed, finite-size glassy domains appear, and the energy of the
resulting QSG
state is reduced accordingly \cite{schechter:06,andresen:13b}. This
leads to a reduction of the energy of the QSG state below that of the
ferromagnetic state, and thus to a zero-temperature phase transition at
finite $H_r$, which is much smaller than $J_0$.  The predicted
functional form for the boundary between the FM and the PM phases at
$T=0$ is given by \cite{andresen:13b}
\begin{equation}
H_r\left(J^{-1}\right) \propto
\left(J^{-1}-J^{-1}_c\right)^{\frac{(3/2-\theta)}{(3-\theta)}}\, .
\end{equation}
Here, $\theta\approx 0.19$ is the stiffness exponent \cite{hartmann:99}.
Indeed, our numerical results are in agreement with this functional
form, as can be seen by fitting the data for $H_r< 1$, fixing the power
to its theoretically obtained value \cite{andresen:13b}
$\beta=(3/2-\theta)/(3-\theta)=0.466$, and using the above estimated
critical disorder $J_c\approx 1.63$ for the fit. This leaves the
proportionality prefactor $\alpha$ as
the single free fitting parameter. The inset of
Fig.~\ref{fig:phase_diagram_h_sigma} shows the fitting result. We note
that our data are limited to $H_r \leq 0.5$. It was not possible to
determine the crossing point of the Binder ratio curves for different
system sizes for random fields with $H_r>0.5$ because of the proximity
of the crossing point to $g=1$.

At finite temperature and for $J \lesssim J_c$, similar considerations
to the ones mentioned above manifest themselves in the dependence of
$T_c$ on the random field. The underlying glassy state at finite
temperature consists of paramagnetic SG domains of typical size $\xi$.
We denote this phase as ``pQSG''. From moderate $H_r$, which does not
affect the typical domain size, the reduction
of the energy per spin of the pQSG state is $\propto H_r /\xi ^{3/2}$.
As a result, only for $H_r > h^* \propto (J_c-J) \xi^{3/2}$ the
disordering of the FM phase is by the pQSG phase. In this regime
theory predicts a linear dependence of $T_c$ on $H_r$ \cite{andresen:13b}.
At $H_r<h^*$ the disordering is to the standard PM
phase, with the known weak dependence of $T_c$ on $H_r$. All the above
considerations do not apply to $J \ll J_c$, where the FM is far from the
SG phase, and the disordering to the PM phase is standard for all
strengths of random field.

In Fig.~\ref{fig:th_phase_diagram} we present the $T$ vs $H_r$ phase
diagrams for disorders $J=0.50$, $J=1.50$, and $J=1.60$. In
Fig.~\ref{fig:th_phase_diagram}(a) $J \ll J_c$. The disordering is into
the standard PM phase, with quadratic dependence $T(H_r)-T_c(0) \propto
H_r^2$ for small $H_r$ and a small reentrance at low temperatures. For
$J=1.50$ [panel (b)] the system is close to the SG phase. Indeed, for
$H_r < h^*\approx 0.1$ the dependence of $T_c$ on $H_r$ is weak, as is
expected for disordering to the standard PM phase. For $H_r > h^*$ we
find the dependence of $T_c$ on $H_r$ to be linear, in agreement with
disordering to the pQSG phase.  We note that at $0.5<H_r<0.6$ there is a
sharp decrease of $T_c$ to zero within the finite-temperature
simulations, and we find a value of $H_r=0.39$ for the zero temperature
transition (inferred from the fitting function in the inset of
Fig.~\ref{fig:phase_diagram_h_sigma}). This is in agreement with the
disordering field by the QSG phase being much smaller than the
interaction strength also at low temperatures. We note that as the QSG
phase is frozen, susceptibility measurements are expected to depict only
the crossover between the QSG and the PM phase at large random fields.
This is in agreement with Ref.~\cite{silevitch:07}, where a sharp
feature in the susceptibility is observed at higher temperatures,
depicting the transition to the pQSG phase, but a smooth crossover is
observed for the lower temperatures. The observed peak value at the
crossover occurs at larger $H_r$ with diminishing peak value as
temperature is reduced \cite{silevitch:07}, in agreement with the
scenario of the QSG having smaller glassy domains as a function of
increasing $H_r$ \cite{schechter:06,andresen:13b}. In panel (c) of
Fig.~\ref{fig:th_phase_diagram} we present the results for $J=1.60$,
closer to the zero-field FM-to-SG transition.  Results are similar to
the case of $J=1.50$, with a weaker linear dependence, and with a
smaller random field, $0.3<H_r<0.4$, which disorders the FM at all
temperatures.

Let us now consider the order of the FM to PM/pQSG transition. The order
of the phase transition of the RFIM is a long standing question. Whereas
analytic arguments support a second-order transition controlled by a
zero-temperature fixed point~\cite{bray:85,villain:85,fisher:86}, some
numerical results support a first-order transition
\cite{young:85,rieger:93} while others support a second-order transition
\cite{ogielski:86,ogielski:86a,rieger:95,ahrens:11,ahrens:13,fytas:13,theodorakis:14}.
For bimodal disorder distributions a mean-field solution suggests a
first-order transition for large enough values of the random
field~\cite{aharony:78a}. Recent numerical work supports universality in
the RFIM~\cite{fytas:13}, suggesting that the nature of the phase
transition is independent of the random-field distribution. Moreover,
high-accuracy estimates for the magnetic exponent ratio
$\beta/\nu$~\cite{theodorakis:14} found the value to be very small, but
clearly finite. In the present work, for small exchange disorder
($J=0.50$) we do not find any signature of a first-order phase
transition. This suggests that the recent phase transition behavior
found for the RFIM~\cite{fytas:13,theodorakis:14} also applies for
models with small exchange disorder.

We now consider the order of the transition in the presence of both
random fields and strong random interactions. Intriguingly, our results
suggest, for strong interaction disorder, a continuous transition for
$H_r<h^*$ where the FM disorders into a ``standard'' PM phase, and a
first-order transition for $H_r>h^*$ where the disordering is into the
pQSG phase. In the latter, we find a discontinuous jump in the
magnetization across the transition, but from our microcanonical
simulations, based on a recently proposed
method~\cite{martin-mayor:07,fernandez:08,fernandez:12a}, we did not
detect any latent heat, which is either zero, or too small to be
detected in the current accessible system sizes, similar to
Refs.~\cite{rieger:93,rieger:95}.

Specifically, we study the distribution of the magnetization above and
below $T_c$ for $J=1.50$ at different field strengths $H_r$, as shown in
Fig.~\ref{fig:magnetization_histogram}. A bimodal distribution close to
$T_c$ is a sign of a first-order phase transition, whereas a normal
distribution with mean $\mu_0=\left[\langle\vert
m\vert\rangle\right]_{\rm av}$ would be the expected distribution for a
continuous phase transition. For small $H_r=0.04<h^*$ [panel (a) of
Fig.~~\ref{fig:magnetization_histogram}] we note that the
single-Gaussian fit and the double-Gaussian fit agree, suggesting that
the magnetization distribution is likely normal close to $T_c$
and therefore the phase transition is continuous. However, we find that
for large $H_r=0.40>h^*$ the double-Gaussian fit does differ
considerably from the single-Gaussian fit, moreover, the double Gaussian
fit reproduces better the magnetization histogram below and above $T_c$.
This suggests that the magnetization jumps at $T_c$, and thus the
existence of a first-order phase transition.

To strengthen the claim of a bimodally-distributed magnetization for
large random fields close to $T_c$, we perform a bootstrap analysis of
the double-Gaussian fit for $J=1.50$ and $H_r=0.30$, $0.40$, and $0.50$,
and show the normalized difference between the mean of the two modes as
a function of the temperature $T$ (see Fig.~\ref{fig:mean_histogram}).
We observe that the maximum values of the curves are close to $T_c$. The
fact that the relative difference between the modes is maximal close to
the critical temperature gives further evidence for a first-order phase
transition.  These distributions show large finite-size corrections and a
clean determination of the critical temperature is difficult.

\section{Discussion}
\label{sec:Discussion}

Disordered nonmagnetic ferroic systems naturally show randomness in both
their inter-particle interactions and in the effective field presented as
a bias between the two degenerate single-particle states.  Recently, it
was shown that this interplay of randomness of interactions and fields
is present also in the ferromagnetic phase of the $\LHx$ compound. This
has allowed for new experiments \cite{silevitch:07,silevitch:10}
investigating this interplay of randomness, as well as for new insights into
older experiments, e.g., demonstrating different final states obtained
by in-field and zero-field annealing of the disordered ferromagnetic
$\LHx$ system \cite{Brooke:99}.

The effective random field in the $\LHx$ system is a consequence of the
interplay between the off-diagonal terms of the dipolar interactions and
the applied transverse field. The applied field also induces transitions
between the spin up $\up$ and spin down $\down$ states in Ho, thus
giving rise to an effective transverse field term in the Ising
Hamiltonian. However, at low field, the effective random field dominates
because it is linear in the applied transverse field
\cite{schechter:06,schechter:08}, whereas the effective transverse field
is negligible at small fields \cite{schechter:05,schechter:08b}.

In a previous study \cite{andresen:13b} we analyzed the $\LHx$
system in the regime where the system is ferromagnetic, albeit with
disorder in both the interaction and the effective longitudinal field.
We have suggested a novel disordering mechanism where at a finite random
field that is much smaller than the typical interaction, finite-size
glassy domains disorder the FM phase into a PM phase. This mechanism
explains various experimental features, such as the linear dependence of
the critical temperature with increased random field, and the
diminishing and rounding of the susceptibility peak with decreasing
temperature \cite{silevitch:07}.

The scaling theory within which the novel disordering mechanism was
obtained \cite{andresen:13b} is not particular to dipolar systems. In
the present work we have shown that the same mechanism applies to a {\em
generic} short-range random-field Ising model with competing
interactions having a ferromagnetic mean. Our results thus support the
generality of the disordering mechanism to random-field ferroic systems
with competing interactions, and it would be of interest to check
their applicability, e.g., to relaxor
ferroelectrics\cite{westphal:92,akbarzadeh:12,sherrington:13,sherrington:15}.

We find an excellent agreement, quantitative and qualitative, between
our numerical results and the predictions of the scaling theory based
on the picture of the disordering of a FM with competing interactions
by a QSG (pQSG) phase at low (high) temperature. It would be of much
interest to further corroborate our results here with a direct microscopic
analysis of the domain structure in the disordered phase.


We further analyze here the nature of the FM to PM transition and show
evidence suggesting that once the disordering is induced via the
above-mentioned mechanism, where the FM phase is disordered by finite
size glassy domains (i.e. to the pQSG phase), then the transition is
first order. This differs from the RFIM with no competing interaction.
In the latter, it is believed that the thermodynamic phase transition is
second order\cite{imry:75}, albeit experimentally hard to be observed
because of slow relaxation.  It would be of much interest to further
study this transition between the FM and pQSG phases, and the dynamics
of the system near the transition, as hysteresis and slow relaxation are
expected. With regard to the latter, nonequilibrium dynamics of the
Ising model were recently studied in the presence of random
bonds\cite{manssen:15} or in the presence of random fields\cite{ohr:17}.
For both systems it was shown that the equilibrium disorder-driven
transition shows up when measuring the nonequilibrium aging properties.
It would be of interest to study whether the quasi-SG driven transition
discussed here has distinct characteristics in the non equilibrium aging
properties of the system.

\begin{acknowledgments}

We would like to thank Amnon Aharony, Michael Moore and David Sherrington for
  useful discussions and valuable input.  J.C.A.~acknowleges support by the
  G\"oran Gustafsson Foundation.  M.S.~acknowledges support from the Israel
  Science Foundation (Grant No.~821/14).  H.G.K.~acknowledges support from the
  National Science Foundation (Grant No.~DMR-1151387) and would like to thank
  Quantum Kellerbier for inspiration. The research of H.G.K.~is based upon work
  supported in part by the Office of the Director of National Intelligence
  (ODNI), Intelligence Advanced Research Projects Activity (IARPA), via MIT
  Lincoln Laboratory Air Force Contract No.~FA8721-05-C-0002. The views and
  conclusions contained herein are those of the authors and should not be
  interpreted as necessarily representing the official policies or endorsements,
  either expressed or implied, of ODNI, IARPA, or the U.S.~Government.  The
  U.S.~Government is authorized to reproduce and distribute reprints for
  Governmental purpose notwithstanding any copyright annotation thereon. We
  thank Texas A\&M University for access to their Ada and Curie clusters, the
  Swedish National Infrastructure for Computing for access to Beskow and
  Triolith clusters, the KTH Department of Theoretical Physics for access to the
  Octopus cluster, Mikael Twenst\"om for access to the Termina cluster, and ETH
  Zurich for access to the Brutus cluster.

\end{acknowledgments}

\section{Appendix}

The simulation parameters for Fig.~\ref{fig:th_phase_diagram} are shown
in Tab.~\ref{table:01}, Tab.~\ref{table:02} and Tab.~\ref{table:03} for
$\tilde J=0.50$, $\tilde J=1.50$, and $\tilde J=1.60$, respectively.
The simulation parameters of the zero-temperature simulations are shown in
Tab.~\ref{table:04}.

\bibliography{refs,comments}

\begin{thebibliography}{72}
\bibitem[{com()}]{comment:aharony}
\bibinfo{note}{{In this work only Ising systems are considered. We note in
  passing that in dilute Heisenberg dipolar magnets ferromagnetism is unstable
  also in the absence of an external field \cite{aharony:78}.}}
\end{thebibliography}

\begin{table*}
\caption{
Parameters of the simulations for $\tilde J=J/J_0=0.5$ where $J_0=1$.
$N_{\rm sa}$ is the number of samples, $N_{\rm sw}$ is the total number
of Monte Carlo sweeps used for equilibration, $T_{\rm min}$ is the
lowest temperature simulated, $T_{\rm max}$ is the highest temperature
simulated, and $N_T$ is the number of temperatures used in the parallel
tempering method for each system size $L$.\label{table:01}}
\begin{tabular*}{2.0\columnwidth}{@{\extracolsep{\fill}} c r r r r r r r}
\hline
\hline
$\tilde J$ & $H_r$ & $L$ & $N_{\rm sa}$ & $N_{\rm sw}$ & $T_{\rm min}$ & $T_{\rm max}$ & $N_{T}$
\\
\hline
$0.50$  & 0.00	& $16,20,24$  & $512$	& $1024$	& $4.200$ & $5.000$ & $100$ \\
$0.50$  & 0.00	& $28,32$  & $512$	& $2048$	& $4.200$ & $5.000$ & $100$ \\
$0.50$  & 0.00	& $40$  & $512$	& $4096$	& $4.200$ & $5.000$ & $100$ \\
$0.50$  & 0.00	& $48$  & $512$	& $8192$	& $4.200$ & $5.000$ & $40$ \\
[1mm]
$0.50$  & 0.20	& $16,20,24$  & $512$	& $2048$	& $4.100$ & $5.000$ & $30$ \\
$0.50$  & 0.20	& $28,32,36,40$  & $512$	& $2048$	& $4.100$ & $5.000$ & $50$ \\
[1mm]
$0.50$  & 0.30	& $16,20,24$  & $1024$	& $2048$	& $4.100$ & $5.000$ & $30$ \\
$0.50$  & 0.30	& $28,32$  & $512$	& $2048$	& $4.100$ & $5.000$ & $50$ \\
$0.50$  & 0.30	& $40$  & $512$	& $4096$	& $4.100$ & $5.000$ & $50$ \\
[1mm]
$0.50$  & 0.50	& $16$  & $1024$	& $4096$	& $4.100$ & $5.000$ & $30$ \\
$0.50$  & 0.50	& $20$  & $1024$	& $8192$	& $4.100$ & $5.000$ & $30$ \\
$0.50$  & 0.50	& $24$  & $1024$	& $16384$	& $4.100$ & $5.000$ & $30$ \\
$0.50$  & 0.50	& $28,32,40$  & $1024$	& $16384$	& $4.100$ & $5.000$ & $50$ \\
[1mm]
$0.50$  & 0.70	& $16$  & $1024$	& $2048$	& $3.950$ & $4.850$ & $30$ \\
$0.50$  & 0.70	& $20$  & $1024$	& $8196$	& $3.950$ & $4.850$ & $30$ \\
$0.50$  & 0.70	& $24$  & $1024$	& $16384$	& $3.950$ & $4.850$ & $30$ \\
$0.50$  & 0.70	& $28,32,40$  & $1024$	& $16384$	& $3.950$ & $4.850$ & $50$ \\
[1mm]
$0.50$  & 0.90	& $12$  & $1024$	& $1024$	& $3.700$ & $4.800$ & $30$ \\
$0.50$  & 0.90	& $16$  & $1024$	& $4096$	& $3.700$ & $4.800$ & $30$ \\
$0.50$  & 0.90	& $20$  & $1024$	& $8192$	& $3.700$ & $4.800$ & $30$ \\
$0.50$  & 0.90	& $24$  & $1024$	& $16384$	& $3.700$ & $4.800$ & $30$ \\
$0.50$  & 0.90	& $28,32$  & $1024$	& $65536$	& $3.700$ & $4.800$ & $40$ \\
[1mm]
$0.50$  & 1.30	& $12$  & $1024$	& $512$	& $2.200$ & $5.000$ & $40$ \\
$0.50$  & 1.30	& $16$  & $1900$	& $32768$	& $2.200$ & $5.000$ & $50$ \\
$0.50$  & 1.30	& $20$  & $1424$	& $262144$	& $2.200$ & $5.000$ & $50$ \\
$0.50$  & 1.30	& $24$  & $1424$	& $1048576$	& $2.200$ & $5.000$ & $50$ \\
[1mm]
$0.50$  & 1.60	& $8$  & $2300$	& $8192$	& $1.500$ & $5.000$ & $50$ \\
$0.50$  & 1.60	& $10$  & $2300$	& $65536$	& $1.500$ & $5.000$ & $50$ \\
$0.50$  & 1.60	& $12$  & $2073$	& $4194304$	& $1.500$ & $5.000$ & $50$ \\
$0.50$  & 1.60	& $14$  & $2116$	& $4194304$	& $1.500$ & $5.000$ & $50$ \\
[1mm]
$0.50$  & 1.80	& $6$  & $1274$	& $32768$	& $1.400$ & $5.000$ & $50$ \\
$0.50$  & 1.80	& $8$  & $2230$	& $65536$	& $1.400$ & $5.000$ & $50$ \\
$0.50$  & 1.80	& $10$  & $2392$	& $2097152$	& $1.400$ & $5.000$ & $50$ \\
$0.50$  & 1.80	& $12$  & $2752$	& $16777216$	& $1.400$ & $5.000$ & $50$ \\
[1mm]
$0.50$  & 2.00	& $4,6$  & $3000$	& $16384$	& $0.600$ & $4.500$ & $50$ \\
$0.50$  & 2.00	& $8$  & $4500$	& $1048576$	& $0.600$ & $4.500$ & $50$ \\
$0.50$  & 2.00	& $10$  & $1200$	& $33554432$	& $0.600$ & $4.500$ & $50$ \\
[1mm]
\hline
\hline
\end{tabular*}
\end{table*}

\begin{table*}
\caption{
Parameters of the simulations for $\tilde J=J/J_0=1.50$, where $J_0=1$.
$N_{\rm sa}$ is the number of samples, $N_{\rm sw}$ is the total number
of Monte Carlo sweeps used for equilibration, $T_{\rm min}$ is the
lowest temperature simulated, $T_{\rm max}$ is the highest temperature
simulated, and $N_T$ is the number of temperatures used in the parallel
tempering method for each system size $L$.  \label{table:02}}
\begin{tabular*}{2.0\columnwidth}{@{\extracolsep{\fill}} c r r r r r r r}
\hline
\hline
$\tilde J$ & $H_r$ & $L$ & $N_{\rm sa}$ & $N_{\rm sw}$ & $T_{\rm min}$ & $T_{\rm max}$ & $N_{T}$
\\
\hline
$1.50$  & 0.00	& $16,20$  & $1024$	& $16384$	& $3.30$ & $3.80$ & $100$ \\
$1.50$  & 0.00	& $24,28$  & $1024$	& $65536$	& $3.30$ & $3.80$ & $100$ \\
$1.50$  & 0.00	& $32,40$  & $1024$	& $262144$	& $3.30$ & $3.80$ & $30$ \\
[1mm]
$1.50$  & 0.02	& $16$	& $1024$	& $65536$ & $3.22$ & $3.983$ & $18$ \\
$1.50$  & 0.02	& $20,24$	&   $1024$	& $65536$ & $3.22$ & $3.983$ & $30$ \\
$1.50$  & 0.02	& $28$	&   $1024$	& $131072$ & $3.22$ & $3.983$ & $30$ \\
$1.50$  & 0.02	& $32,40$	&   $1024$	& $131072$ & $3.22$ & $3.983$ & $40$ \\
[1mm]
$1.50$  & 0.04	& $16$  & $1024$	& $65536$	& $3.22$ & $3.983$ & $18$ \\
$1.50$  & 0.04	& $20$  & $1024$	& $65536$	& $3.22$ & $3.983$ & $40$ \\
$1.50$  & 0.04	& $24,28,32,40$  & $1024$	& $131072$	& $3.22$ & $3.983$ & $40$ \\
[1mm]
$1.50$  & 0.06	& $16$  & $1024$	& $32768$	& $3.22$ & $3.983$ & $18$ \\
$1.50$  & 0.06	& $20,24$  &   $1024$	& $65536$	& $3.22$ & $3.983$ & $40$ \\
$1.50$  & 0.06	& $28,32$  &   $1024$	& $262144$	& $3.22$ & $3.983$ & $40$ \\
$1.50$  & 0.06	& $40$  &   $1200$	& $524288$	& $3.22$ & $3.983$ & $40$ \\
[1mm]
$1.50$  & 0.08	& $16,20$  & $1024$	& $16384$	& $3.25$ & $4.00$ & $20$ \\
$1.50$  & 0.08	& $24$  & $1024$	& $65536$	& $3.22$ & $4.00$ & $18$ \\
$1.50$  & 0.08	& $28$  & $1024$	& $262144$	& $3.22$ & $4.00$ & $40$ \\
$1.50$  & 0.08	& $32,40$  & $1024$	& $524288$	& $3.22$ & $4.00$ & $40$ \\
[1mm]
$1.50$  & 0.10	& $16$  & $2048$	& $16384$	& $3.20$ & $4.00$ & $16$ \\
$1.50$  & 0.10	& $20$  & $2048$	& $32768$	& $3.20$ & $4.00$ & $16$ \\
$1.50$  & 0.10	& $24$  & $1024$	& $65536$	& $3.20$ & $4.00$ & $16$ \\
$1.50$  & 0.10	& $28$  & $1024$	& $262144$	& $3.20$ & $4.00$ & $40$ \\
$1.50$  & 0.10	& $32,40$  & $1024$	& $524288$	& $3.20$ & $4.00$ & $40$ \\
[1mm]
$1.50$  & 0.20	& $12$  & $2048$	& $16384$	& $3.00$ & $4.00$ & $16$ \\
$1.50$  & 0.20	& $16$  & $1024$	& $32768$	& $3.00$ & $4.00$ & $16$ \\
$1.50$  & 0.20	& $20,24$  & $1024$	& $131072$	& $3.00$ & $4.00$ & $16$ \\
$1.50$  & 0.20	& $28,32$  & $1024$	& $524288$	& $3.00$ & $4.00$ & $30$ \\
[1mm]
$1.50$  & 0.30	& $12,14$  & $2048$	& $32768$	& $2.78$ & $4.00$ & $20$ \\
$1.50$  & 0.30	& $16$  & $1024$	& $65536$	& $2.78$ & $4.00$ & $20$ \\
$1.50$  & 0.30	& $20$  & $2043$	& $131072$	& $2.78$ & $4.00$ & $20$ \\
$1.50$  & 0.30	& $24$  & $3285$	& $1048576$	& $2.78$ & $4.00$ & $30$ \\
[1mm]
$1.50$  & 0.40	& $8,10,12$  & $2048$	& $32768$	& $2.00$ & $4.50$ & $60$ \\
$1.50$  & 0.40	& $14,16$  & $2048$	& $65536$	& $2.00$ & $4.50$ & $60$ \\
$1.50$  & 0.40	& $20$  & $1024$	& $1048576$	& $2.00$ & $4.50$ & $60$ \\
[1mm]
$1.50$  & 0.50	& $4$  & $4096$	& $32768$	& $1.00$ & $5.00$ & $50$ \\
$1.50$  & 0.50	& $6$  & $3637$	& $65536$	& $1.00$ & $5.00$ & $50$ \\
$1.50$  & 0.50	& $8$  & $2048$	& $131072$	& $1.00$ & $5.00$ & $50$ \\
$1.50$  & 0.50	& $10$  & $2048$	& $262144$	& $1.00$ & $5.00$ & $50$ \\
$1.50$  & 0.50	& $12$  & $1441$	& $1048576$	& $1.00$ & $5.00$ & $50$ \\
$1.50$  & 0.50	& $14$  & $2645$	& $2097152$	& $1.00$ & $5.00$ & $50$ \\
[1mm]
$1.50$  & 0.60	& $4,6$  & $2048$	& $16384$	& $1.00$ & $5.00$ & $50$ \\
$1.50$  & 0.60	& $8$  & $1387$	& $65536$	& $1.00$ & $5.00$ & $50$ \\
$1.50$  & 0.60	& $10$  & $2048$	& $262144$	& $1.00$ & $5.00$ & $50$ \\
$1.50$  & 0.60	& $12$  & $5275$	& $524288$	& $1.00$ & $5.00$ & $50$ \\
$1.50$  & 0.60	& $14$  & $1199$	& $4194304$	& $1.00$ & $5.00$ & $50$ \\
$1.50$  & 0.60	& $16$  & $1024$	& $33554432$	& $1.00$ & $5.00$ & $50$ \\
[1mm]
\hline
\hline
\end{tabular*}
\end{table*}

\begin{table*}
\caption{
Parameters of the simulations for $\tilde J=J/J_0=1.60$, where $J_0=1$. $N_{\rm
sa}$ is the number of samples, $N_{\rm sw}$ is the total number of Monte Carlo
sweeps used for equilibration, $T_{\rm min}$ is the lowest temperature
simulated, $T_{\rm max}$ is the highest temperature simulated, and $N_T$ is the
number of temperatures used in the parallel tempering method for each system
size $L$.  \label{table:03}}
\begin{tabular*}{2.0\columnwidth}{@{\extracolsep{\fill}} c r r r r r r r}
\hline
\hline
$\tilde J$ & $H_r$ & $L$ & $N_{\rm sa}$ & $N_{\rm sw}$ & $T_{\rm min}$ & $T_{\rm max}$ & $N_{T}$
\\
\hline
$1.60$  & 0.00	& $12$  & $1024$	& $8192$	& $2.75$ & $5.00$ & $100$ \\
$1.60$  & 0.00	& $16$  & $1024$	& $32768$	& $2.75$ & $5.00$ & $100$ \\
$1.60$  & 0.00	& $20$  & $1024$	& $65536$	& $2.75$ & $5.00$ & $100$ \\
$1.60$  & 0.00	& $24$  & $1024$	& $262144$	& $2.75$ & $5.00$ & $30$ \\
$1.60$  & 0.00	& $28$  & $1024$	& $524288$	& $2.75$ & $5.00$ & $30$ \\
$1.60$  & 0.00	& $32$  & $1024$	& $1048576$	& $2.75$ & $5.00$ & $30$ \\
[1mm]
$1.60$  & 0.10	& $12$  & $1024$	& $32768$	& $1.75$ & $5.00$ & $37$ \\
$1.60$  & 0.10	& $16,20$  & $1024$	& $131072$	& $1.75$ & $5.00$ & $37$ \\
$1.60$  & 0.10	& $24$  & $1024$	& $262144$	& $2.75$ & $5.00$ & $30$ \\
$1.60$  & 0.10	& $28$  & $1024$	& $524288$	& $2.75$ & $5.00$ & $30$ \\
$1.60$  & 0.10	& $32$  & $1024$	& $1048576$	& $2.75$ & $5.00$ & $30$ \\
[1mm]
$1.60$  & 0.20	& $8$  & $2048$	& $8192$	& $1.75$ & $5.00$ & $37$ \\
$1.60$  & 0.20	& $10$  & $2840$	& $16384$	& $1.75$ & $5.00$ & $37$ \\
$1.60$  & 0.20	& $12$  & $1500$	& $65536$	& $1.75$ & $5.00$ & $37$ \\
$1.60$  & 0.20	& $14$  & $2048$	& $1048576$	& $1.75$ & $5.00$ & $40$ \\
$1.60$  & 0.20	& $16$  & $2048$	& $4194304$	& $1.75$ & $5.00$ & $40$ \\
[1mm]
$1.60$  & 0.30	& $4$  & $3024$	& $8192$	& $0.5$ & $4.00$ & $71$ \\
$1.60$  & 0.30	& $6$  & $3024$	& $32768$	& $0.5$ & $4.00$ & $71$ \\
$1.60$  & 0.30	& $8$  & $4096$	& $131072$	& $0.5$ & $4.00$ & $71$ \\
$1.60$  & 0.30	& $10$  & $2440$	& $1048576$	& $0.5$ & $4.00$ & $71$ \\
$1.60$  & 0.30	& $12$  & $1782$	& $8388608$	& $0.5$ & $4.00$ & $71$ \\
[1mm]
$1.60$  & 0.40	& $4$  & $4000$	& $8192$	& $0.50$ & $4.00$ & $71$ \\
$1.60$  & 0.40	& $6$  & $2048$	& $32768$	& $0.50$ & $4.00$ & $71$ \\
$1.60$  & 0.40	& $8$  & $5800$	& $131072$	& $0.50$ & $4.00$ & $71$ \\
$1.60$  & 0.40	& $10$  & $3500$	& $524288$	& $0.50$ & $4.00$ & $71$ \\
$1.60$  & 0.40	& $12$  & $2100$	& $4194304$	& $0.50$ & $4.00$ & $71$ \\
\hline
\hline
\end{tabular*}
\end{table*}

\begin{table*}
\caption{
Parameters of the zero-temperature simulations with $\tilde J=J/J_0$, where
$J_0=1$. $H_r$ is the random field strength, $N_{\rm sa}$ is the number of
samples, $N_{\rm eo}$ is the total number of simulation steps for each system
size $L$.
\label{table:04}}
\begin{tabular*}{2.0\columnwidth}{@{\extracolsep{\fill}} c l r r r }
\hline
\hline
$\tilde J$ & $H_r$ & $L$ & $N_{\rm sa}$ & $N_{\rm eo}$
\\
\hline
$0.00$  &$1.40$, $1.50$, $1.55$, $1.60$, $1.70$, $1.80$	& $4$  & $1024$	& $10000000$ \\
$0.00$  &$1.40$, $1.50$, $1.55$, $1.60$, $1.65$, $1.70$, $1.75$, $1.80$	& $6$  & $1024$	& $25000000$ \\
$0.00$  &$1.40$, $1.50$, $1.55$, $1.60$, $1.65$, $1.68$, $1.70$, $1.75$	& $8$  & $1024$	& $35000000$ \\
$0.00$  &$1.50$, $1.55$, $1.60$, $1.70$			& $10$  & $1024$& $268435456$ \\
[1mm]
$0.15$  &$1.50$, $1.60$, $1.70$				& $4$  & $1024$	& $15000000$ \\
$0.15$  &$1.50$, $1.55$, $1.60$, $1.65$, $1.70$		& $6$  & $1024$	& $25000000$ \\
$0.15$  &$1.50$, $1.55$, $1.60$, $1.65$, $1.70$  	& $8$  & $1024$	& $35000000$ \\
[1mm]
$0.25$  &$1.50$, $1.55$, $1.60$, $1.70$	    & $4$  & $1024$	&  $25000000$ \\
$0.25$  &$1.40$, $1.50$, $1.55$, $1.60$, $1.65$, $1.70$  & $6$  & $1024$ & $25000000$ \\
$0.25$  &$1.40$, $1.50$, $1.55$, $1.60$, $1.65$		& $8$  & $1024$	& $35000000$ \\
[1mm]
$0.50$  &$1.40$, $1.50$, $1.55$, $1.60$, $1.65$		& $4$  & $1024$	& $25000000$ \\
$0.50$  &$1.40$, $1.45$, $1.50$, $1.55$, $1.60$		& $6$  & $1024$	& $35000000$ \\
$0.50$  &$1.40$, $1.45$, $1.50$, $1.55$, $1.60$		& $8$  & $1024$	& $35000000$ \\
\hline
\hline
\end{tabular*}
\end{table*}

\end{document}